# Exploring thermodynamic and photonic properties of new black hole solutions in $F(R)$ gravity theory

M. Dehghani[1*]

1- *Department of Physics, Razi University, Kermanshah, Iran*

## Abstract

In this paper, we investigate black hole solutions in $F(R)$ gravity within a four-dimensional spacetime, separately treating the cases of vanishing and non-vanishing energy-momentum tensors. We demonstrate that in the absence of an energy-momentum tensor, the number of independent equations is sufficient to obtain the Ricci-flat solutions without the need for a prescribed $F(R)$ function. In the same case, also the solutions with $r$- dependent Ricci scalar can obtained too, but the number of independent equations is insufficient, necessitating an appropriate $R(r)$ ansatz. Furthermore, by incorporating Maxwell's electromagnetic theory to account for a non-vanishing energy-momentum tensor, we derive exact solutions for both constant and $r$-dependent Ricci scalars, thereby introducing new charged black holes. Utilizing this new method, we not only recover the general relativistic and previously known $F(R)$ gravity solutions but also introduce four new classes of $F(R)$ black holes. We calculate their thermodynamic quantities and confirm the validity of the first law of black hole thermodynamics. The thermal stability of these new black holes is analyzed using both thermodynamic and geometrical approaches. Finally, the photon spheres and black hole shadows are studied within the $F(R)$ gravity framework.

## 1- Introduction

Einstein's theory of General Relativity (GR), which describes gravity as the curvature of space-time geometry, has provided an extraordinary framework for understanding gravitational phenomena across various scales. This theory has successfully withstood rigorous classic tests, such as the precession of Mercury's perihelion and the deflection of light by intense gravitational fields. Furthermore, the direct detection of gravitational waves by LIGO-Virgo collaborations has provided robust experimental confirmation of GR [1, 2, 3].

However, on cosmological scales in extreme gravitational environments, GR faces significant theoretical and observational challenges. On a cosmological scale, the standard framework of GR necessitates the introduction of dark energy and dark matter to account for the observed accelerated expansion of the Universe. On the other hand, in the high-curvature regimes near the centers of black holes, GR is inevitably leads to the formation of singularities, the points where space-time curvature and energy density become infinite [4]. Such singularities are physically problematic and indicate that GR may not a complete fundamental theory. This suggests that GR must be extended or modified in these regimes, potentially pointing toward a more comprehensive theory, possibly a theory of quantum gravity [4].

[*]e-mail: m.dehghani@razi.ac.ir

To address these shortcomings, modified gravity theories emerged as a prominent research direction. One of the most well-studied generalizations is $F(R)$ gravity, where the standard Einstein-Hilbert action is replaced by an arbitrary function of the Ricci scalar $F(R)$ [5, 6]. This modification introduces new degrees of freedom into the gravitational field, offering a powerful tool to investigate the mitigation or removal of central singularities [7, 8, 9]. Moreover, $F(R)$ gravity can describe diverse gravitational phenomena, from inflationary era of the early Universe to late-time cosmic acceleration, without the need to invoke unknown fields [5, 10, 11].

Various specific forms of $F(R)$ gravity have been extensively studied. For instance, the power-law model $F(R) \sim R^N$ has proven successful in explaining both cosmic inflation and the late-time acceleration of the Universe [12, 13]. Similarly, the Strobinsky model $F(R) = R + \alpha R^2$, along with other prominent models [14, 15, 16, 17], provides a robust framework for understanding cosmic evolution. Beyond cosmology, the search for exact solutions in $F(R)$ gravity has been a centeral theme, leading to the discovery of new charged, spherically symmetric black holes [18, 19, 20], as well as regular and multi-horizon black hole configurations when coupled with nonlinear electrodynamics [21]. More recently, the study of photon sphere and black hole shadow has demonstrated that the results of $F(R)$ gravity are highly consistent with the observations of M87* and Sgr A* [22].

While much research has focused on cosmological implications, the derivation of exact analytical solutions in four-dimensional $F(R)$ gravity remains a fundamental challenge. Most existing studies rely on a ''top-down'' approach, where a specific functional form of $F(R)$ is pre-selected as an ansatz, and the field equations are subsequently solved [23, 24, 25, 26]. In contrast, in the case of constant Ricci scalar, which is frequently assumed in the black hole literature, the independent equations are sufficient to determine all unknowns without requiring an $F(R)$ ansutz [27].

In this work, we propose a distinct and more flexible "bottom-up" methodology. Rather than assuming a special $F(R)$ function a priori, our approach recognizes that all unknowns can be determined directly in the constant Ricci scalar case, while offering a much suitable framework in the more general case of variable curvature $R(r)$. We demonstrate that by treating the Ricci scalar as an $r$-dependent function, one can derive the appropriate $F(R)$ model for a given $R(r)$ profile. Using this novel approach, we introduce new classes of exact charged and uncharged black hole solutions and rigorously investigate their thermodynamics, stability, and optical properties, including photon spheres and black hole shadows.

This paper is organized as follows: In section 2 the formalism of four-dimensional $\mathrm{F}(R)$ gravity and the resulting field equations are explicitly derived. Section 3 is devoted to solving these equations in two general cases including zero and non-zero energy-momentum tensors. Through this approach, four new classes of charged and uncharged exact solutions are introduced. It is shown that our solutions recover the general relativistic-based Schwarzschild–anti-de Sitter (S-AdS) and Reissner-Nordström-anti-de Sitter (R-N-AdS) black holes, as well as the previously known $\mathrm{F}(R)$ gravity exact solutions. In section 4, the geometrical properties of these solutions are explored. The existence of event horizons, physical singularities and the asymptotic AdS behaviors are confirmed via curvature scalars analysis. Section 5 presents the calculation of thermodynamic quantities and demonstration of the correctness of the first law of thermodynamics. In section 6 the thermal stability and phase transitions of the black holes are investigated using both the canonical ensemble and geometrical methods. Finally, section 7 examines the optical characteristics of these black holes by analyzing null geodesics, providing exact expressions for the photon sphere radii and the black hole shadows. The results are summarized and discussed in section 8.

## 2- The field equations

The action of four-dimensional $\mathrm{F}(R)$ black holes in the presence of cosmological constant can be written as [28]

$$I=-\frac{1}{16\pi}\int d^4x\sqrt{-g}[F(R)-2\Lambda+L(\mathcal{F})] \tag{2.1}$$

Where, $R=g^{\mu\nu}\mathcal{R}_{\mu\nu}$ is the Ricci scalar, $\Lambda=-3\ell^{-2}$ is the four-dimensional AdS cosmological constant and, $L(\mathcal{F})$ with $\mathcal{F}=F_{\mu\nu}F^{\mu\nu}$ is the Lagrangian of the electromagnetic theory under consideration. The anti-symmetric tensor $F_{\mu\nu}$ in terms of electromagnetic four-potential $A_\mu$ is defined as $F_{\mu\nu}=\partial_\mu A_\nu-\partial_\nu A_\mu$.
By use of the variational principle, we obtain the gravitational and electromagnetic field equations as

$$F_R R_{\mu\nu}-\frac{1}{2}F g_{\mu\nu}+\left(g_{\mu\nu}\nabla_\alpha\nabla^\alpha-\nabla_\mu\nabla_\nu\right)F_R+\Lambda g_{\mu\nu}=T_{\mu\nu}, \tag{2.2}$$

$$\nabla_\mu[L_{\mathcal{F}}(\mathcal{F})F^{\mu\nu}]=0. \tag{2.3}$$

where, we have used the notations $F=F(R)$, $F_{\mathrm{R}}=\frac{dF}{dR}$ and $L_{\mathcal{F}}(\mathcal{F})=\frac{dL(\mathcal{F})}{d\mathcal{F}}$, for simplicity. The general form of the electromagnetic energy-momentum tensor takes the following form

$$T_{\mu\nu}=\frac{1}{2}g_{\mu\nu}L(\mathcal{F})-2\,L_{\mathcal{F}}(\mathcal{F})F_{\mu\alpha}F_\nu^\alpha. \tag{2.4}$$

Here, we are interested in solving the field equations in a four-dimensional spherically symmetric geometry identified by the following line element [36]

$$ds^2=-B(r)dt^2+\frac{dr^2}{B(r)}++r^2\left(d\theta^2+\sin^2\theta\,d\varphi^2\right). \tag{2.5}$$

In which, $B(r)$ is an unknown function of the radial coordinate $r$, known as the metric function. It will be determined later. In this geometry, the explicit forms of the $tt$, $rr$, $\theta\theta$ and $\varphi\varphi$ components of the gravitational field equation (2.2) are obtained as

$$C_{tt} = \left(B'' + \frac{2B\prime}{r}\right)F_R - \left(B' + \frac{4B}{r}\right)F_R' - 2BF_R'' + F - 2\Lambda = \frac{2}{B}T_{tt}, \quad (2.6)$$

$$C_{rr} = \left(B'' + \frac{2B\prime}{r}\right)F_R - \left(B' + \frac{4B}{r}\right)F_R' + F - 2\Lambda = -2BT_{rr}, \quad (2.7)$$

$$C_{\theta\theta} = (1 - B - rB')F_R + \mathrm{r}(B + rB')F_R' + r^2BF_R'' - \left(\frac{F}{2} - \Lambda\right)r^2 = T_{\theta\theta}, \quad (2.8)$$

$$C_{\varphi\varphi} = (1 - B - rB')F_R + \mathrm{r}(B + rB')F_R' + r^2BF_R'' - \left(\frac{F}{2} - \Lambda\right)r^2 = \frac{T_{\varphi\varphi}}{sin^2\theta}. \quad (2.9)$$

Also, by taking the trace of Eq. (2.2) one obtains

$$T_{race} = RF_R - 2F + 4\Lambda + 3\left(B' + \frac{2B}{r}\right)F_R' + 3BF_R'' = T, \quad (2.10)$$

Where, $T = \mathrm{g}^{\mu\nu}T_{\mu\nu}$ is trace of the energy-momentum tensor. Moreover, by use of the line element (2.5), after some tensor calculations, we obtained

$$R = g^{\mu\nu}R_{\mu\nu} = -B'' - \frac{4B'}{r} - \frac{2(B-1)}{r^2}. \quad (2.11)$$

Note that in overall this paper prime means derivative with respect to $r$.

## 3- The exact solutions

Now, we obtain exact solutions of the equations obtained in the previous section for two general separated cases

### 3-1- Zero energy-momentum tensor ($T_{\mu\nu} = 0$)

Noting Eqs. (2.6)-(2.10), we have $F_R'' = 0$, and

$$C_{tt} = C_{rr} = F - 2\Lambda + \left(B'' + \frac{2B\prime}{r}\right)F_R - \left(B' + \frac{4B}{r}\right)F_R' = 0, \quad (3.1)$$

$$C_{\theta\theta} = C_{\varphi\varphi} = (1 - B - rB')F_R + r(B + rB')F_R' - \left(\frac{F}{2} - \Lambda\right)r^2 = 0, \quad (3.2)$$

$$T_{race} = RF_R - 2F + 4\Lambda + 3\left(B' + \frac{2B}{r}\right)F_R' = 0. \quad (3.3)$$

Now, we proceed by considering the following important cases:

#### 3-1-1- The solutions with constant $R$ ($r$-independent)

Since $R$ is treated as a constant, we have $F_R' = 0$, too. Therefore, we have the following more simplified equations

$$F - 2\Lambda + \left(B'' + \frac{2B'}{r}\right) F_R = 0, \tag{3.4}$$

$$(1 - B - rB')F_R - \left(\frac{F}{2} - \Lambda\right) r^2 = 0, \tag{3.5}$$

$$RF_R - 2F + 4\Lambda = 0. \tag{3.6}$$

In terms of the constant $\alpha$ with the dimension of $Length^{-1}$, the Eq.(3.6) can be solved to give $F(R)$ in the following form

$$F(R) = \alpha R^2 + 2\Lambda. \tag{3.7}$$

Now, through combining Eqs. (3.5) and (3.6), we have

$$B' + \frac{B-1}{r} + \frac{R}{4} r = 0. \tag{3.8}$$

The solution of this first-order differential equation, in terms of the integration constant $-m$, gives the metric function $B(r)$ in the following form

$$B(r) = 1 - \frac{m}{r} - \frac{R}{12} r^2. \tag{3.9}$$

For the case $R = 4\Lambda$, (3.9) recovers the 4D S-AdS black hole. Generally speaking, we have introduced a class of new $F(R)$ black hole identified by Eqs. (3.7) and (3.9). It is easy to show that these solutions satisfy Eq. (3.4) too.

## 3-1-2- The solutions with variable $R$ ($r$-dependent)

In this case the relation $F_R'' = 0$, is again satisfied. Its solution can be written as

$$F_R = F_0 + F_1 r, \tag{3.10}$$

$$F(r) = \int R'(F_0 + F_1 r) dr + F_2, \tag{3.11}$$

Where, $F_0$, $F_1$ and $F_2$ are constants of integration and, $R' = \frac{dR(r)}{dr}$. Noting Eqs. (3.1)-(3.3), by simple calculations, one can show that

$$\frac{2C_{\theta\theta}}{r^2} - C_{tt} = T_{ract}. \tag{3.12}$$

It means that only two of three equations (3.1)-(3.3) are independent. Therefore, we have three unknowns ($F, R$ and B) and two independent equations. Thus this theory is faced to the problem of under determinacy. In this case we have to guess one of the unknowns. Thus, we take $R$ in the following form

$$R(r) = 4\Lambda + \frac{c}{r^2}\,, \tag{3.13}$$

and, $c$ is a dimensionless constant. Return to (2.11) and solving the related differential equation, we have

$$B(r) = \frac{2-c}{2} + \frac{C_1}{r} + \frac{C_2}{r^2} - \frac{1}{3}\Lambda\, r^2. \tag{3.14}$$

Where, $C_1$ and $C_2$ are integration constant. Combining (3.13) and (3.11), gives

$$F(r) = F_2 + \frac{2cF_1}{r} + \frac{cF_0}{r^2}. \tag{3.15}$$

Note that $F_0$ is dimensionless, $F_1$ and $F_2$ have dimensions of $Length^{-1}$and $Length^{-2}$, respectively. With the aim of fixing the constants, we put the values of $R(r)$, $B(r)$ and $F(r)$ into Eqs. (3.1)-(3.3). It gives the following constraints

$$F_1(c-1) = 0, \qquad F_2 = 2\Lambda(F_0+1), \qquad F_0 C_2 = 0, \qquad cF_0 = 3F_1C_{1.} \tag{3.16}$$

At this stage the following distinct cases are considerable

- $c = 1$ and $F_0 \neq 0$, $F_1 \neq 0$,

Thus, we have $F_2 = 2\Lambda(F_0+1), C_1 = F_0/(3F_1),$ and $C_2 = 0,$

$$R(r) = 4\Lambda + \frac{1}{r^2}, \tag{3.17}$$

$$B(r) = \frac{1}{2} + \frac{F_0}{3F_1 r} - \frac{1}{3}\Lambda\, r^2, \tag{3.18}$$

$$F(r) = 2\Lambda(\mathrm{F}_0+1) + \frac{2F_1}{r} + \frac{F_0}{r^2}. \tag{3.19}$$

By choosing $F_0 = 1$ and $\frac{1}{3\mathrm{F}_1} = -m$, the results can be summarized as

$$B(r) = \frac{1}{2} - \frac{m}{r} - \frac{1}{3}\Lambda\, r^2, \tag{3.20}$$

$$F(R) = R \pm \frac{2}{3\mathrm{m}}\sqrt{R - 4\Lambda}, \tag{3.21}$$

Which are just the same as obtained in ref. [20, 29].

- $c = 1$, $C_2 \neq 0$ and $F_1 \neq 0$,

In this case, we have $\mathrm{F}_0 = 0$, $\mathrm{F}_2 = 2\Lambda$, $\mathrm{C}_2 = -m\, r_0$ ($r_0$ is a dimensional constant), and $\mathrm{C}_1 = 0$. Therefore,

$$R(r) = 4\Lambda + \frac{1}{r^2}, \tag{3.22}$$

$$B(r) = \frac{1}{2} - \frac{m\, r_0}{r^2} - \frac{1}{3}\Lambda\, r^2, \tag{3.23}$$

$$F(R) = 2\Lambda \pm 2F_1\sqrt{R - 4\Lambda}, \tag{3.24}$$

- $c = 0$ and $F_0 \neq 0$,

We have $\mathrm{F}_1 = 0$, $\mathrm{C}_2 = 0$, and $R = 4\Lambda$. Thus

$$B(r) = 1 + \frac{\mathrm{C}_1}{r} - \frac{1}{3}\Lambda r^2, \tag{3.25}$$

$$F(R) = 2\Lambda(F_0 + 1). \tag{3.26}$$

By choosing $F_0 = 1$, and $\mathrm{C}_1 = -m$ it is just the SAdS black holes of GR.

## 3-2- Non-zero energy-momentum tensor ($T_{\mu\nu} \neq 0$)

For obtaining the charged $F(R)$ black holes it is necessary to obtain the simultaneous solutions of (2.3) and (2.6)-(2.10). By use of the Maxwell's electrodynamics, in terms of the integration constant $q$, the only nonzero component of the Faraday's tensor is $F_{tr} = \frac{q}{r^2}$. Then the components of the energy-momentum tensor $T_{\mu\nu}$, take the following forms

$$T_{\varphi\varphi} = \frac{q^2}{r^2} sin^2\theta, \qquad T_{\theta\theta} = \frac{q^2}{r^2}, \qquad T_{rr} = \frac{-q^2}{Br^4}, \qquad T_{tt} = \frac{Bq^2}{r^4}, \tag{3.27}$$

After replacing in Eqs. (2.6)-(2.10) and noting the relations $F_R'' = 0$, and $T = 0$, we have

$$C_{tt} = C_{rr} = \left(B'' + \frac{2B'}{r}\right)F_R - \left(B' + \frac{4B}{r}\right)F_R' + F - 2\Lambda = \frac{2q^2}{r^4}, \tag{3.28}$$

$$C_{\theta\theta} = C_{\varphi\varphi} = (1 - B - rB')F_R + \mathrm{r}(B + rB')F_R' - \left(\frac{F}{2} - \Lambda\right)r^2 = \frac{q^2}{r^2}, \tag{3.29}$$

$$T_{race} = RF_R - 2F + 4\Lambda + 3\left(B' + \frac{2B}{r}\right)F_R' = 0. \tag{3.30}$$

Now, we solve these differential equations in the following distinct cases.

### 3-2-1- The solutions with constant $R$ ($r$-independent)

In this special case, we have $F_R' = 0$ and Eqs. (3.28)-(3.30) will be simplified such that the Eq. (3.12) is fulfilled. Then by solving the two remaining independent equations, the unknowns $F(R)$ and $B(r)$ are determined as

$$F(R) = \alpha R^2 + 2\Lambda, \tag{3.31}$$

$$B(r) = 1 - \frac{m}{r} + \frac{q^2}{r^2} - \frac{R}{12}\, r^2. \tag{3.32}$$

Here, $\alpha$ and $m$ are the constants of integration, and the dimensionless constant $F_R$ has been absorbed in $q^2$.

### 3-2-2- The solutions with variable $R$ ($r$-dependent)

In the case of $r$-dependent Ricci scalar, it is possible to show that there are three unknowns, while the number of independent equations is two. To overcome this problem, we proceed by guessing $R(r)$ in the following form

$$R(r) = 4\Lambda + \frac{c}{r^2} + \frac{b}{r^5}, \tag{3.33}$$

Where, $c$ and $b$ are two constants. By solving the differential equation (2.11), in terms of the integration constants $\mathrm{C}_1$ and $\mathrm{C}_2$, we have

$$B(r) = \frac{2-c}{2} - \frac{b}{2r^3} + \frac{\mathrm{C}_1}{r} + \frac{\mathrm{C}_2}{r^2} - \frac{\Lambda}{3} r^2. \tag{3.34}$$

Then, by solving the equation $F_R'' = 0$, through the relation (3.11), we have

$$F(r) = F_2 + \frac{bF_0}{r^5} + \frac{5bF_1}{4r^4} + \frac{cF_0}{r^2} + \frac{2cF_1}{r}. \tag{3.35}$$

First of all, by letting $c = 0$, and examining the Eqs. (3.33)-(3.35) into Eqs. (3.28)-(3.30), we have $b = 0, \mathrm{F1} = 0,\ C_2 = q^2$, and $F_2 = 2\Lambda(F_0 + 1)$. Now, by choosing $F_0 = 1$, and $C_1 = -m$, we have

$$B(r) = 1 - \frac{m}{r} + \frac{q^2}{r^2} - \frac{\Lambda}{3} r^2, \qquad \text{and} \qquad F(R) = R = 4\Lambda, \tag{3.36}$$

which are nothing but just the R-N-AdS solutions.

In order to fix the constants $c, b, F_0, F_1, F_2, C_1$ and $C_2$, for the general case, we substitute Eqs. (3.33)-(3.35) into (3.28)-(3.30). That gives

$$F_0 = 0, \quad F_1(c-1) = 0, \quad F_2 = 2\Lambda(F_0 + 1), \quad 8F_0C_2 = 8q^2 + 5bF_1, \quad cF_0 = 3F_1C_1. \tag{3.37}$$

Now we consider the case corresponding to $c = 1$ and $F_1 \neq 0$. In this case, we have $\mathrm{F}_2 = 2\Lambda, 5\mathrm{bF}_1 = -8\mathrm{q}^2, \mathrm{C}_1 = 0$, and the results are as follows

$$R(r) = 4\Lambda + \frac{1}{r^2} - \frac{8\mathrm{q}^2}{5\mathrm{F}_1 r^5}, \tag{3.38}$$

$$B(r) = \frac{1}{2} + \frac{4\mathrm{q}^2}{5\mathrm{F}_1 r^3} + \frac{\mathrm{C}_2}{r^2} - \frac{\Lambda}{3} r^2, \tag{3.39}$$

$$F(r) = 2\Lambda - \frac{2q^2}{r^4} + \frac{2F_1}{r}. \tag{3.40}$$

By introducing the new constants $\frac{4}{5F_1} = r_0$, and $C_2 = -m\, r_0$ ($r_0$ is a dimensional constant), Eqs. (3.38)-(3.40) can be rewtitten as

$$R(r) = 4\Lambda + \frac{1}{r^2} - \frac{2r_0 q^2}{r^5}, \tag{3.41}$$

$$B(r) = \frac{1}{2} + \frac{r_0 q^2}{r^3} - \frac{m\, r_0}{r^2} - \frac{\Lambda}{3} r^2, \tag{3.42}$$

$$F(r) = 2\Lambda - \frac{2q^2}{r^4} + \frac{8}{5 r r_0}. \tag{3.43}$$

Therefore, we were able to introduce two new sets of charged $F(R)$ black holes as presented in Eqs. (3.31) and (3.32) for constant Ricci scalar and, Eqs. (3.41)-(3.43) for $r$-dependent Ricci scalar.

## 4- Geometrical properties of the solutions

In the previous section, we have introduced four sets of new exact solutions. They have presented in Eqs. (3.7) and (3.9), Eqs. (3.22)-(3.24), Eqs. (3.31) and (3.32), and Eqs. (3.41)-(3.43). Additionally, our method is capable of producing S-AdS black holes [Eqs. (3.25) and (3.26)] and, R-N-AdS black holes [Eq. (3.36)] as well as the previously obtained $F(R)$ solutions [Eqs. (3.17), (3.20) and (3.21)].

At this stage, we discuss existence of the spacetime singularities and asymptotic behavior of the solutions by calculating the Ricci ($R = g^{\mu\nu} R_{\mu\nu}$) and Kretschmann ($K = R^{\mu\nu\alpha\beta} R_{\mu\nu\alpha\beta}$) scalars.

- $$B(r) = 1 - \frac{m}{r} - \frac{R}{12} r^2 \tag{4.1}$$
  $$R = R_0 = \text{Constant}, \qquad K = \frac{12m^2}{r^6} + \frac{R^2}{6}. \tag{4.2}$$

- $$B(r) = \frac{1}{2} - \frac{m\, r_0}{r^2} - \frac{\Lambda}{3} r^2 \tag{4.3}$$
  $$R = \frac{1}{r^2} + 4\Lambda, \qquad K = \frac{56 m^2 r_0^2}{r^8} + \frac{4 m r_0}{r^6} + \frac{1}{r^4} + \frac{4\Lambda}{3r^2} + \frac{8\Lambda^2}{3}. \tag{4.4}$$

- $$B(r) = 1 - \frac{m}{r} + \frac{q^2}{r^2} - \frac{R}{12} r^2, \tag{4.5}$$
  $$R = R_0 = \text{Constant}, \qquad K = \frac{56 q^4}{r^8} - \frac{48\, m\, q^2}{r^7} + \frac{12\, m^2}{r^6} + \frac{R^2}{6}. \tag{4.6}$$

- $$B(r) = \frac{1}{2} - \frac{m r_0}{r^2} + \frac{q^2 r_0}{r^3} - \frac{\Lambda}{3} r^2 \tag{4.7}$$
  $$R = \frac{1}{r^2} - \frac{2q^2 r_0}{r^5} + 4\Lambda, \tag{4.8}$$

$$K = \frac{1}{r^4} - \frac{4q^2 r_0}{r^7} + \frac{4m\, r_0}{r^6} + \frac{184 q^4 r_0^2}{r^{10}} - \frac{200 m q^2 r_0^2}{r^9} + \frac{56 m^2 r_0^2}{r^8} + \frac{4\Lambda}{3r^2} - \frac{8q^2 r_0 \Lambda}{3r^5} + \frac{8\Lambda^2}{3}. \tag{4.9}$$

Now, by taking the limits, we have

$$\lim_{r\to 0} R = \infty, \qquad \text{and} \qquad \lim_{r\to 0} K = \infty. \tag{4.10}$$

It mean that there is a physical singularity at the origin which is covered by the event horizon. Also, one can show that

$$\lim_{r\to \infty} R = 4\Lambda \ \text{ or } \ R_0, \qquad \text{and} \qquad \lim_{r\to \infty} K = \frac{8\Lambda^2}{3} \ \text{or} \ \frac{{R_0}^2}{6}, \tag{4.11}$$

Which confirm that our solutions are asymptotically A(dS). Additionally, it is clear from the metric functions of Eqs. (4.1), (4.3), (4.5), and (4.7) that at least there is one real root for the equation $B(r) = 0$. All above discussions prove that our solutions pass two criteria for indicating black holes: existing a physical singularity and possessing at least one event horizon. In the next section, we explore thermodynamic properties of these black holes.

## 5- Thermodynamics

With the aim of satisfaction of the first law of black hole thermodynamics, we proceed to calculate the thermodynamic and conserved quantities of the novel $F(R)$ black holes introduced here.

At first, we notice that the black hole temperature associated with the black hole horizon can be calculated by utilizing the concept of surface gravity. In terms of the surface gravity $\kappa$, we have $T = \kappa/(2\pi)$ and $\kappa = \sqrt{-(\nabla_\mu \chi_\nu)(\nabla^\mu \chi^\nu)/2}$. Taking $\chi^\mu = (-1, 0, 0, 0)$, one can show that [30]

$$T = \frac{1}{4\pi} B'(r_+), \tag{5.1}$$

and with the given metric function $B(r)$ it can be calculated. Here, $r_+$ is the outermost event horizon which is obtained by imposing the condition $B(r_+) = 0$.

The black hole entropy, as a crucial thermodynamic quantity, can be calculated by use of the entropy-area law. Based on this nearly universal law, the entropy of $F(R)$ black holes is equal to one-fourth of the horizon surface area, multiplied by $F_R$. Therefore, we have

$$S = \pi r_+^2 F_R(r_+), \tag{5.2}$$

The other thermodynamic quantity which is required to be calculated is the electric potential associated with the black hole horizon. It can be calculated with respect to a reference point located at a large distance from the black hole horizon. Thus, we can use the following standard relation [31, 32]

$$U(r_+) = C\mathrm{A}_\mu \chi^\mu\big|_{refrence} - C\mathrm{A}_\mu \chi^\mu\big|_{r=r_+} = C\mathrm{A}_t(r_+). \tag{5.3}$$

It means that with the temporal component of four-potential in hand, one can calculate the electric potential on the horizon. The coefficient constant $C$ will be fixed later.

Now, we extend the studies for the new black holes introduced here.

- $B(r) = 1 - \frac{m}{r} - \frac{R}{12}r^2, \qquad F(R) = \alpha R^2 + 2\Lambda, \qquad R = R_0$ =Constant. (5.4)

By using (5.1), the temperature on the horizon is obtained as

$$T = \frac{1}{4\pi r_+}\left(1 - \frac{R}{4}r_+^2\right). \tag{5.5}$$

Assuming the black hole mass $M$, is a constant multiple of $m$ (i.e. $M = k\, m$), and calculating $m$ by imposing the condition $B(r_+) = 0$, we have

$$M = k\, r_+\left(1 - \frac{R}{12}r_+^2\right), \tag{5.6}$$

Now, regarding Eq. (5.2), one can chow that $\partial M/\partial S = T$, provided that $k_1$is fixed to $k = F_R/2$. In that case we have $M = F_R m/2$ and $dM = TdS$, which is just the first law of black hole thermodynamics.

- $B(r) = \frac{1}{2} - \frac{m\, r_0}{r^2} - \frac{\Lambda}{3}r^2, \qquad F(R) = 2\Lambda + 2F_1\sqrt{R - 4\Lambda}, \qquad R = \frac{1}{r^2} + 4\Lambda.$ (5.7)

For the temperature and entropy, we have

$$T = \frac{1}{4\pi r_+}\left(1 - \frac{4}{3}\Lambda r_+^2\right), \qquad S = \pi r_+^2 F_R(r_+). \tag{5.8}$$

The black hole's mass can be written as

$$M = \frac{k r_+^2}{r_0}\left(r_+ - \frac{4}{3}\Lambda\, r_+^3\right). \tag{5.9}$$

Now, it can be shown that $\partial M/\partial S = T$, provided that $k = 3/4$, and $F_1 r_0 = 1$, are chosen. Under these conditions the first law is valid in the form $dM = TdS$.

- $B(r) = 1 - \frac{m}{r} + \frac{q^2}{r^2} - \frac{R}{12}r^2, \qquad F(R) = \alpha R^2 + 2\Lambda.$ (5.10)

$$T = \frac{1}{4\pi}\left(\frac{1}{r_+} - \frac{q^2}{r_+^3} - \frac{R}{4}r_+\right), \qquad S = \pi r_+^2 F_R. \tag{5.11}$$

$$M = k\left(r_+ + \frac{q^2}{r_+} - \frac{R}{12}r_+^3\right). \tag{5.12}$$

By use of Eq. (5.3), the electric potential is calculated as

$$U(r_+) = C\mathrm{A}_t(r_+) = \frac{Cq}{r_+}. \tag{5.13}$$

Noting the fact that the black hole's electric charge $Q$, we have $Q = q$, and making use of Eqs. (5.14), (5.15), and (5.16), one can show that $(\partial M/\partial S)_Q = T$ and $(\partial M/\partial Q)_S = U$, provided that that the constants fixed to $k = F_R/2$, and $C = 1$. Therefore, under these conditions the first law is valid in the following form

$$dM = TdS + \mathrm{U}dQ. \tag{5.14}$$

- $$B(r) = \frac{1}{2} - \frac{mr_0}{r^2} + \frac{q^2 r_0}{r^3} - \frac{\Lambda}{3} r^2 \tag{5.15}$$

$$R(r) = \frac{1}{r^2} - \frac{2q^2 \mathrm{r}_0}{r^5} + 4\Lambda, \qquad F(r) = 2\Lambda - \frac{2\mathrm{q}^2}{r^4} + \frac{8}{5rr_0}. \tag{5.16}$$

$$T = \frac{1}{4\pi r_+}\left(1 - \frac{q^2 r_0}{r_+^3} - \frac{4}{3}\Lambda\, r_+^2\right), \qquad S = \pi r_+^2 F_R. \tag{5.17}$$

$$M = \frac{k}{r_0}\left(\frac{r_+^2}{2} + \frac{q^2 r_0}{r_+} - \frac{1}{3}\Lambda\, r_+^4\right). \tag{5.18}$$

By use of Eq. (5.3), the electric potential is given by Eq. (5.13).

Noting the fact that the black hole's electric charge $Q$, we have $Q = q$, and making use of Eqs. (5.17), (5.18), and (5.13), one can show that $(\partial M/\partial S)_Q = T$ and $(\partial M/\partial Q)_S = U$, provided that the constants fixed to $k_1 = 3/5$, and $C = 6/5$. Therefore, under these conditions the first law is valid in the form of Eq. (5.14).

## 6- Stability

In this section, we analyze the black hole thermal stability by use of the canonical ensemble and geometrical method, comparatively. In the canonical ensemble, the points of type-1 and type-2 phase transitions and the local stability of the black holes can be studied regarding the signature of the black hole heat capacity. In the geometrical method, the type-1 and type-2 phase transitions are characterized by the divergent points of the Ricci scalar of a proposed thermodynamic metric. Here, we investigate the local stability or phase transitions of the new $F(R)$ black holes introduced in the present work.

## 6-1- Local stability in the canonical ensemble

To investigate the local stability of the charged black holes and identify the locations of type-1 and type-2 phase transitions, it is essential to evaluate the heat capacity. At a constant $Q$, the heat capacity is defined by the following thermodynamic relation [33]

$$C_Q = T\left(\frac{\partial S}{\partial T}\right)_Q = \frac{T}{M_{SS}}, \qquad \text{with} \qquad M_{SS} = \left(\frac{\partial^2 M}{\partial S^2}\right)_Q. \tag{6.1}$$

Within the canonical ensemble framework, the local stability of a charged black hole is dictated by the sign of its heat capacity. Positivity of heat capacity signifies a local stable configuration. Conversely, unstable black holes undergo either type-1 or type-2 phase transition to reach a stable state. These transitions are characterized by the vanishing and diverging points of heat capacity, respectively [34, 35].

Following the derivation of the temperature in the preceding section, we proceed to compute the heat capacity. For the charged black hole solutions described by the metric function $B(r)$, the second derivative of the mass is given by:

- $B(r) = 1 - \frac{m}{r} + \frac{q^2}{F_R\, r^2} - \frac{R}{12} r^2, \qquad R = R_0. \qquad (6.2)$

$$M_{SS} = \frac{-1}{8\pi^2 r_+ F_R}\left(\frac{1}{r_+} + \frac{R}{4} - \frac{3q^2}{F_R r_+^4}\right), \tag{6.3}$$

- $B(r) = \frac{1}{2} - \frac{m r_0}{r^2} + \frac{q^2 r_0}{r^3} - \frac{\Lambda}{3} r^2, \qquad (6.4)$

$$M_{SS} = \frac{-1}{12\pi^2 r_+ F_R}\left(\frac{1}{r_+^2} + \frac{4}{3}\Lambda - \frac{4q^2 r_0}{r_+^5}\right), \tag{6.5}$$

with $R(r)$ and $F(r)$ are defined in Eq. (5.16).

The thermal behavior of these black holes is illustrated in figure 1, where the temperature $T$ (blue curve) and the heat capacity $C_Q$(black curve) are plotted against the horizon radius $r_+$. In the left panel, we observe two type-1 phase transition points, denoted as $r_{1ext}$ and $r_{2ext}$, along with a type-2 phase transition at $r_+ = r_1$. The analysis indicates that black holes within the interval $r_{1ext} < r_+ < r_1$ are locally stable. The right panel further corroborates the existence of one type-1 transition at $r_+ = r_{ext}$ and one type-2 phase transitions at $r_+ = r$, confirming that stability is maintained in the region $r_{ext} < r_+ < r$.

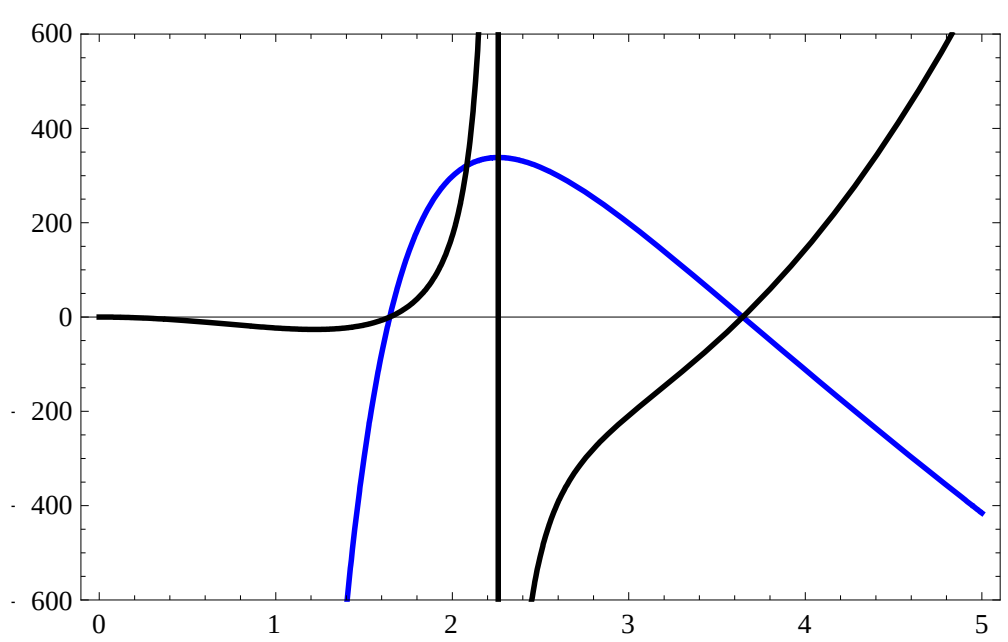

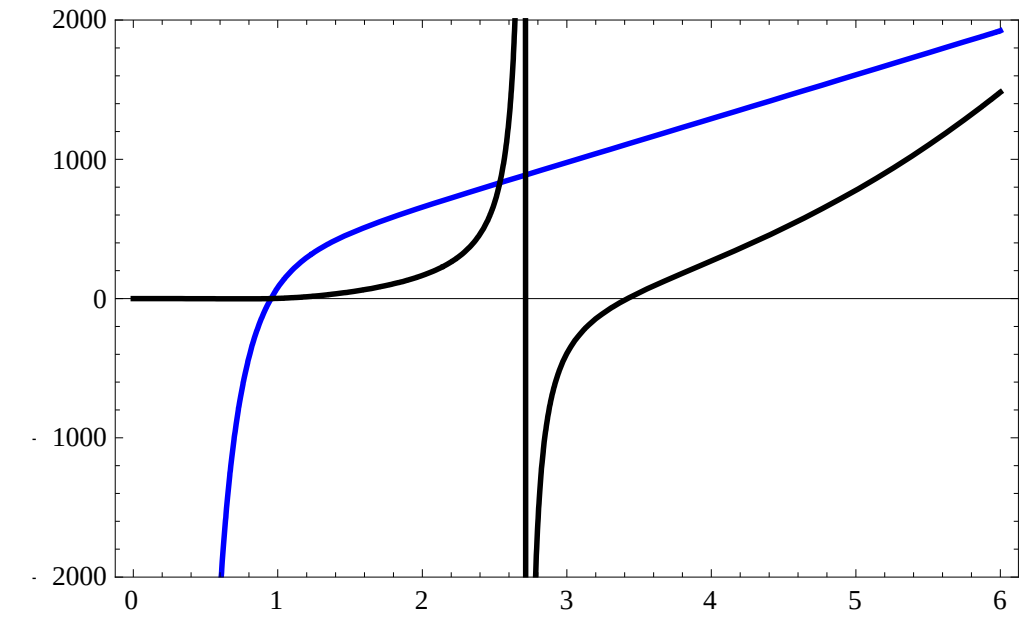

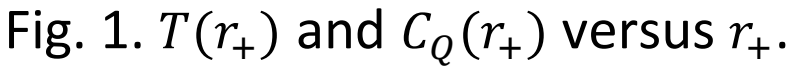

Fig. 1. $T(r_+)$ and $C_Q(r_+)$ versus $r_+$.

Left: $C_Q$, $40000\,T$ for $R = 0.25, F_R = 2, q = 1.5$. Right: $C_Q$, $1000\,T$ for $q = 2, r_0 = 1, \Lambda = -3$.

## 6-2- Geometrical thermodynamics

Beyond the standard canonical ensemble approach, various frameworks have been developed to investigate the thermodynamic phase transition of the black holes. Among these, geometrical thermodynamics has emerged as a prominent method, where the phase transition points are encoded within the curvature of a specifically constructed thermodynamic metric [36, 37]. The divergence of the Ricci scalar associated with these metrics serves as a key indicators of phase transitions. Several well-known thermodynamic metrics have been proposed in the literature, most notably those by Quevedo [38], Weinhold [39], Ruppeiner [40], and the recently introduced HPEM (Hendi, Panahiyan, Eslampanah, and Momennia) metric [41]. Here, we define the relevant metric structure as follows:
The two models proposed by Quevedo (QI and QII) are given by

$$ds^2 = \Omega\left(-M_{SS}ds^2 + M_{QQ}dQ^2\right), \quad \text{with} \quad \Omega = \begin{cases} SM_S + QM_Q, & \text{for} \quad (\mathrm{Q}I), \\ SM_S\,, & \text{for} \quad (\mathrm{Q}II). \end{cases} \tag{6.6}$$

Furthermore, the Weinhold ($W$) and Ruppeiner ($R$) metrics are expressed as

$$ds^2 = Mg_{ab}^W dX^a dX^b, \quad \text{with} \quad g_{ab}^W = \frac{\partial^2 M}{\partial X^a \partial X^b}, \tag{6.7}$$

$$ds^2 = -MT^{-1}g_{ab}^W dX^a dX^b. \tag{6.8}$$

The HPEM metric is defined as

$$ds^2 = S\frac{M_S}{M_{QQ}^3}\left(-M_{SS}dS^2 + M_{QQ}dQ^2\right). \tag{6.9}$$

In the geometrical thermodynamics framework, the location of phase transitions correspond to the singular points of the Ricci scalar $\mathcal{R}$, which are determined by the roots of its denominator. Now, we identify the denominators ($D$) for each metric [42]

$$D\left(\mathcal{R}^{(QI)}\right) = 2(M_{SS})^2(M_{QQ})^2(SM_S + QM_Q)^3, \tag{4.13}$$

$$D\left(\mathcal{R}^{(QII)}\right) = 2S^3(M_{SS})^2(M_{QQ})^2(M_S)^3, \tag{4.14}$$

$$D\left(\mathcal{R}^{(R)}\right) = TM^2[(M_{SS}M_{QQ} - (M_{SQ})^2]^2, \tag{4.15}$$

$$D\left(\mathcal{R}^{(W)}\right) = M^2[(M_{SS}M_{QQ} - (M_{SQ})^2]^2, \tag{4.16}$$

$$D\left(\mathcal{R}^{(HPEM)}\right) = 2S^3(M_{SS})^2(M_S)^3. \tag{4.17}$$

In our case, the black hole mass $M$ is a function of two independent variables $Q$ and $S$. The subscripts in $M$ denote the partial derivatives with respect to these variables. Based on the vanishing of the denominators, we evaluate the compatibility of these metrics with the canonical ensemble results:

● The singularities of $\mathcal{R}^{(QI)}$ occur at $M_{SS} = 0$ and $SM_S + QM_Q = 0$. However, these conditions yield results that are inconsistent with the canonical ensemble method.

● Given that $M_{QQ} \neq 0$, $\mathcal{R}^{(QII)}$diverges at $M_S = 0$ and $M_{SS} = 0$. Notably, these points coincide with the phase transition locations identified via the canonical ensemble.

● The phase transition points of Ruppeiner metric are determined $T = 0$, $M = 0$, and $M_{SS}M_{QQ} - (M_{SQ})^2 = 0$. In general, these do not align with the canonical ensemble predictions.

● The Weinhold metric identifies phase transitions at $M = 0$ and $M_{SS}M_{QQ} - (M_{SQ})^2 = 0$, which also shows a discrepancy with the results of canonical ensemble method.

● From Eq.(4.17), it is evident that the points corresponding to type-one and type-two phase transitions, as identified via the HPEM metric, are determined by the roots of $M_S = 0$ and $M_{SS} = 0$. These results are in full agreement with the findings from the canonical ensemble method.

Consequently, our analysis demonstrate that for the newly introduced black holes solutions, the Quevedo II and HPEM metrics provide a consistent description of the thermodynamic phase transitions, yielding results that are perfectly aligned with the canonical ensemble method.

## 7- Photon sphere and black hole shadow

The circular orbit of the photon around a black hole is called a photon circle or, in aggregation, a photon sphere. With the radius of photon sphere in hand, the radius of the black hole shadow can be calculated, easily. Here, we examine the radii of the photon sphere and the black hole shadow of our new black holes obtained in the framework of $F(R)$ modified gravity. To this end we use the following Lagrangian [22]

$$\mathcal{L} = \frac{1}{2} g_{\mu\nu}\dot{q}^{\mu}\dot{q}^{\nu} = \frac{1}{2}\left[-B(r)\dot{t}^2 + \frac{\dot{r}^2}{B(r)} + r^2\dot{\theta}^2 + r^2\sin^2\theta\,\dot{\varphi}^2\right]. \tag{7.1}$$

Here dot means derivative with respect to the affine parameter. Evidently, for the massless photons we have $\mathcal{L} = 0$. For simplicity we choose $\theta = \pi/2$, and $\dot{\theta} = 0$. Therefore, from Eq. (7.1), we have

$$\mathcal{L} = \frac{1}{2} g_{\mu\nu}\dot{q}^{\mu}\dot{q}^{\nu} = \frac{1}{2}\left[-B(r)\dot{t}^2 + \frac{\dot{r}^2}{B(r)} + r^2\dot{\varphi}^2\right]. \tag{7.2}$$

The momentum conjugate to the coordinate $q$ is defined as $p_i = \partial\mathcal{L}/\partial\dot{q}_i$. Since the Lagrangian is independent of $t$ and $\varphi$, the corresponding conjugate momenta are conserve. Therefor

$$p_t = \partial\mathcal{L}/\partial\dot{t} = -B(r)\,\dot{t} = E. \qquad \text{and} \qquad p_\varphi = \partial\mathcal{L}/\partial\dot{\varphi} = r^2\,\dot{\varphi} = L. \tag{7.3}$$

Where, $E$ and $L$ are known as the constants of the motion. The total energy $\mathcal{E}$ of the system can be calculated as [22]

$$\mathcal{E} = \sum_i p_i \dot{q}_i - \mathcal{L} = \mathcal{L}. \tag{7.4}$$

Now, by using Eqs. (7.2) and (7.3), the total conserved energy which is equal to zero can be written as

$$0 = -\frac{E^2}{2} + \frac{\dot{r}^2}{2} + \frac{L^2}{2r^2} B(r). \tag{7.5}$$

By imposing the condition of circular orbits $\dot{r} = 0$, we find the effective potential $V_{eff}$, in the following form

$$V_{eff}(r) = \frac{L^2}{2r^2} B(r) - \frac{E^2}{2}. \tag{7.6}$$

By using the condition $dV_{eff}/dr = 0$, one can calculate the radius $r_{ph}$ of photon circle through solving the following relation

$$rB'(r) - 2B(r) = 0. \tag{7.7}$$

Then, the radius $r_{sh}$ of black hole shadow is obtained making use of the relation

$$r_{sh} = \frac{r_{ph}}{\sqrt{B(r_{ph})}}\,. \tag{7.8}$$

At this stage, by using Eqs. (7.7) and (7.8), we calculate the radii of the photon sphere and the black hole shadow for all the $F(R)$ black holes introduced here.

- $$B(r) = 1 - \frac{m}{r} - \frac{R}{12} r^2 \tag{7.9}$$

  $$r_{ph} = \frac{3m}{2}, \qquad r_{sh} = \frac{3\sqrt{3}m}{2\sqrt{1 - \frac{9R}{16}m^2}}, \qquad 9Rm^2 < 16. \tag{7.10}$$

- $$B(r) = \frac{1}{2} - \frac{m\, r_0}{r^2} - \frac{\Lambda}{3} r^2 \tag{7.11}$$

  $$r_{ph} = 2\sqrt{m\, r_0}, \qquad r_{sh} = \frac{4\sqrt{m r_0}}{\sqrt{1 - \frac{16\Lambda}{3} m r_0}}, \qquad 16\Lambda m r_0 < 3. \tag{7.12}$$

- $$B(r) = 1 - \frac{m}{r} + \frac{q^2}{r^2} - \frac{R}{12} r^2, \qquad \frac{3m}{r_{ph}} - 2 = \frac{4q^2}{r_{ph}^2}, \tag{7.13}$$

  $$r_{ph} = \frac{3m}{4}\left(1 + \sqrt{1 - \frac{32q^2}{9m^2}}\right), \qquad r_{sh} = \frac{r_{ph}}{\sqrt{\frac{1}{2} - \frac{m}{4r_{ph}} - \frac{R}{12} r_{ph}^2}}, \qquad \frac{1}{2} > \frac{m}{4r_{ph}} + \frac{R}{12} r_{ph}^2, \tag{7.14}$$

  which, in the case of $q = 0$, are compatible with those of Eq. (7.10).

- $$B(r) = \frac{1}{2} - \frac{m r_0}{r^2} + \frac{q^2 r_0}{r^3} - \frac{\Lambda}{3} r^2, \qquad \frac{4m r_0}{r_{ph}^2} - 1 = \frac{5q^2 r_0}{r_{ph}^3} \to r_{ph}^3 - 4m r_0 r_{ph} + 5q^2 r_0 = 0. \tag{7.15}$$

Now, Eq. (7.15) can be written as $r_{ph}^3 - a\, r_{ph} + b = 0$, with $a = 4m\mathrm{r}_0$ and $b = 5q^2\mathrm{r}_0$. It can possess three real roots as [43]

$$r_{ph} = 2\sqrt{\frac{a}{3}}\, \sin\left[\frac{1}{3}\sin^{-1}\left(\frac{3b}{2}\sqrt{\frac{3}{a^3}}\right) - \frac{2n\pi}{3}\right], \qquad n = 0, 1, 2, \tag{7.16}$$

which in the case, $n = 0, q = 0$ (or $b = 0$), with $\sin^{-1}(0) = \pi$, reduces to that of Eq. (7.12).

Then, using Eq. (7.8), the radius of black hole shadow can be calculated as

$$r_{sh} = \frac{r_{ph}}{\sqrt{\frac{3}{10} - \frac{mr0}{5r_{ph}^2} - \frac{\Lambda}{3}r_{ph}^2}}, \qquad \frac{3}{10} > \frac{mr0}{5r_{ph}^2} + \frac{\Lambda}{3}r_{ph}^2. \tag{7.17}$$

The radii of photon spheres and black hole shadows presented in Eqs. (7.10), (7.12), (7.14), (7.16) and (7.17) are remarkably consistent with those calculated for the S-AdS and R-N-AdS black holes [44].

## 8- Conclusions

In this work, we have investigated the exact charged and uncharged black hole solutions within the framework of four-dimensional $F(R)$ gravity. By deriving the explicit field equations in a spherically symmetric geometry, we demonstrated that for a constant Ricci scalar (i.e. $R = R_0$), the number of independent field equations is sufficient to determine all unknown functions without requiring a prescribed $F(R)$ form. In the case of an $r$-dependent $R(r)$, we addressed the inherent mathematical under-determination by employing a suitable $R(r)$ ansatz. Consequently, we have introduced four new sets of exact solutions with the following explicit forms:

- $B(r) = 1 - \frac{m}{r} - \frac{R}{12}r^2, \qquad F(R) = \alpha R^2 + 2\Lambda, \qquad R = R_0 =$Constant, (8.1)
- $B(r) = \frac{1}{2} - \frac{m\,\mathrm{r}_0}{r^2} - \frac{\Lambda}{3}r^2, \qquad F(R) = 2\Lambda + 2F_1\sqrt{R - 4\Lambda}, \qquad R = \frac{1}{r^2} + 4\Lambda,$ (8.2)
- $B(r) = 1 - \frac{m}{r} + \frac{q^2}{r^2} - \frac{R}{12}r^2, \qquad F(R) = \alpha R^2 + 2\Lambda, \qquad R = R_0 =$Constant, (8.3)
- $B(r) = \frac{1}{2} - \frac{mr_0}{r^2} + \frac{q^2r_0}{r^3} - \frac{\Lambda}{3}r^2, \qquad F(r) = 2\Lambda - \frac{2\mathrm{q}^2}{r^4} + \frac{8}{5rr_0}, \qquad R(r) = \frac{1}{r^2} - \frac{2q^2\mathrm{r}_0}{r^5} + 4\Lambda.$ (8.4)

Notably, our approach not only recovers the standard general relativistic solutions, specifically the S-AdS and R-N-AdS black holes, but also reproduces previously known $F(R)$ gravity solutions [20]. Through curvature scalars analysis, we confirmed that these solutions possess a physical singularity at the origin, which is shielded by an event horizon, thereby satisfying the fundamental criteria for black hole solutions. Also, the behavior of the curvature scalars at infinity, demonstrate AdS asymptotes of the black holes.

After calculating the curvature scalars and taking the limits, we showed that the new exact solutions include a physical singularity at the origin which is covered by event horizon. Noting this property and also existence of the real roots for $B(r) = 0$, we concluded that our solutions are really black holes.

Furthermore, we evaluated the thermodynamic quantities including entropy, temperature, electric charge, potential, and mass. The validity of the first law of black hole thermodynamics was verified for all newly derived $F(R)$ solutions. The thermal stability of these black holes was investigated using both canonical ensemble and geometrical methods. A comparative analysis reveals that certain stability regimes exist, and specifically, the results QII and HPEM show full compatibility with the canonical ensemble method.

Finally, we examined the optical characteristics of these black holes by analyzing null geodesics. Utilizing the Euler-Lagrange formalism, we obtained exact expressions for the photon sphere radii and the $F(R)$ black holes shadows. The results show that the optical characteristics of our $F(R)$ solutions are remarkably consistent with those of the GR-based S-AdS and R-N-AdS black holes. These findings suggest that future research could extend this work by incorporating nonlinear electrodynamics, potentially leading to discovery of non-singular and multi-horizon $F(R)$ black holes.